# Improved Stability Analysis of Nonlinear Networked Control Systems over Multiple Communication Links

Rahim Delavar,   Babak Tavassoli,   Mohammad Taghi Hamidi Beheshti,

*Abstract*— In this paper, we consider a nonlinear networked control system (NCS) in which controllers, sensors and actuators are connected via several communication links. In each link, networking effects such as the transmission delay, packet loss, sampling jitter and data packet miss-ordering are captured by time-varying delays. Stability analysis is carried out based on the Lyapunov Krasovskii method to obtain a condition for stability of the nonlinear NCS in the form of linear matrix inequality (LMI). The results are applied to a two degrees of freedom robot arm NCS which shows a considerable improvement with respect to the previous works.

## I. INTRODUCTION

There is a noticeable increase in applications of the networked communication for implementation of control systems. Such a control system is referred to as a networked control system (NCS). Due to the several advantages of networking, such as lower cost, less wiring and more reliability, a considerable attention has been drawn to the NCS's in recent years [1]. Usage of the communication network in the control loop, can cause problems due to the several effects of the communication such as:

- Transmission, queuing and processing delays.
- Medium access constraints (waiting times).
- Data packet loss due to either transmission errors or network congestion.
- Data packet miss-ordering.
- Jitter (time-variation) in sampling.

The above networking phenomena reduce the control performance and may even cause system instability [1, 2]. Extensive research has been conducted to study the effects of networking on the stability and performance of the NCS. Most of them only consider one or two of the above phenomena and only a limited number of results can be applied to analyze several of them simultaneously. To mention some of these works, the impact of packet loss has been studied in [3, 4], delay and packet loss in [5, 6, 7] and communication constraints in [8]. The impacts of time-varying transmission intervals, variable time delays, and access constraints are simultaneously considered in [9] where exponential stability and $L_2$-gain analyses are carried out and also in [2] where $L_p$ stability of the nonlinear NCS is investigated based on a hybrid system modeling. There are other works that allow for limited nonlinearity in the form of a perturbation term such as [14]. In [10], the case of nonlinear NCS is considered and an approach is proposed for applying the Lyapunov–Krasovskii method which cannot be directly applied to a nonlinear NCS. It is shown that improvements are achieved with respect to the analysis in [2].

In this work, we consider a nonlinear NCS with different types of delay, data packet loss, sampling jitter and miss-ordering that are captured by a lumped-delay element in the same way as

.

in [10]. We improve the stability analysis in [10] by expanding the calculations in terms of the state norm into the calculations in terms of the absolute values of the state variables. The result is a new sufficient condition for the stability of the nonlinear NCS. This result is applied to obtained a bound on the control cycle of the robot arm NCS introduced in [10] where a comparison shows that the new stability analysis is much better.

*Notation*: The set of real numbers is denoted by $\mathbb{R}$. For $a \in \mathbb{R}$ its absolute value is denoted by $|a|$. For a vector $x \in \mathbb{R}^n$ we denote the $i$th element of $x$ by $[x]_i$. We also define $\bar{x}$ to be the element-wise absolute value of the vector $x$ such that $[\bar{x}]_i = |[x]_i|$. Similarly, for a matrix $A \in \mathbb{R}^{n \times n}$ we denote its element at $i$th row and $j$th column by $[A]_{ij}$ and define $\bar{A}$ such that $[\bar{A}]_{ij} = |[A]_{ij}|$. In a symmetric matrix, one of the every two off-diagonal elements or blocks that have symmetry may be replaced by an asterisk symbol $*$ for brievity.

## II. Network Induced Delay

In this work, a nonlinear networked control system is considered in which the measured values are asynchronously sampled and transmitted over multiple communication links. The effects of communication in each link (transmission delay, packet loss and sampling jitter) are captured by time-varying delay elements. As in [10], the combination of plant, controllers and communication links, can be represented as interconnection of an augmented system and multiple delay channels as in the block diagram in Fig.1 where $q$ is the number of the delay channels. It is mentioned that a data packet loss in communication networks can be also considered as a delay equal to one sampling interval until receiving the next data sample (this idea is applied for example in [11]).

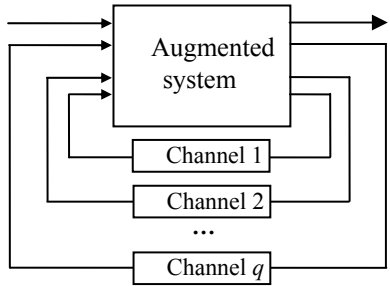

Figure 1. NCS with multiple network-induced delay channels.

In this work, the system in Figure 1, is represented by a nonlinear state space model with time varying delays

$$\dot{x} = f(t, x, x_{d_1}, x_{d_2}, \ldots, x_{d_q}) \tag{1}$$

in which $t \in \mathbb{R}$ is the continuous time, $x(t) \in \mathbb{R}^{n_x}$ is the state of the augmented system, $d_i(t) \in \mathbb{R}$, $1 \leq i \leq q$ is the time-varying delay in the $i$th delay channel. Also, for the sake of compactness of the equations $x(t - d_i(t))$ for $1 \leq i \leq q$ is briefly written as $x_{d_i}(t)$ and the dependencies on $t$ are omitted whenever no ambiguity occurs. As mentioned, there are several communication phenomena that can be captured by a time-varying delay element. An efficient way of avoiding complexities caused by these phenomena is to make the following assumption.

*Assumption 1:* for every $i \in \{1, ..., q\}$, the $i$th delay $d_i$, is allowed to vary uncertainly with time $t$ provided that there exists a positive constant $r_i$ such that $d_i \leq r_i$

It is not difficult to obtain the bounds $r_i$. For example if the sum of different types of delay (queuing, medium access, processing and transmission) for each data packet is bounded by $\eta_m$, the sampling intervals are bounded by $h$ and the number of successive packet losses is bounded by $n$, then $d(t) < \eta_m + nh$.

*Remark 1*: In the non-networked case in which there are no communication delays we have $d_i(t) = 0$ for $1 \leq i \leq q$ and the system (1) is reduced to $\dot{x} = f(t, x, ..., x)$ which will be referred to as the non-networked system.

## III. MAIN RESULTS

In this Section stability of the system described by (1) is studied. For this purpose, we provide sufficient stability conditions in terms of the bounds $r_i$ on delays $d_i(t)$, $1 \leq i \leq q$ in Assumption 1. The analysis is based on the Lyapunov-Krasovskii method in [12]. In order to derive the stability condition, we need to make a series of assumptions on the structure of the system in (1). First, the controller must be designed such that closed loop system, in the non-networked case, is asymptotically stable as in the following.

*Assumption 2:* There exist a continuously differentiable function $V_1(t, x)$ and matrices $S, \bar{F}, \bar{W} \in \mathbb{R}^{n \times n}$ such that the relations in (2), (3) and (4) are satisfied.

$$\left[\frac{\partial V_1}{\partial t} + \frac{\partial V_1}{\partial x} f(t, x, ..., x)\right] \leq -\bar{x}^T S \bar{x}. \tag{2}$$

$$\overline{\left(\frac{\partial V_1}{\partial x}\right)} \leq 2\bar{x}^T W^T. \tag{3}$$

$$\overline{f(t, x, x, ..., x)} \leq \bar{F}\bar{x}. \tag{4}$$

We also make the following assumption about the function $f$ in (1).

*Assumption 3:* There exist matrices $M_k \in \mathbb{R}^{n \times n}$, $1 \leq k \leq q$, such that function $f$ in (1) satisfies the following relations for every $1 \leq k \leq q$.

$$\overline{[\Gamma_k - \Gamma_{k+1}]} \leq M_k \overline{(x - x_{d_k})}. \tag{5a}$$

$$\Gamma_k = f\left(t, x, ..., x, x_{d_k}, x_{d_{k+1}}, ..., x_{d_{q+1}}\right). \tag{5b}$$

$$\Gamma_{q+1} = f(t, x, x, ..., x). \tag{5c}$$

In the above relations, the dependency of $\Gamma_k$ on the variables $t, x, x_{d_k}, x_{d_{k+1}}, ..., x_{d_{q+1}}$ is not shown for brevity.

*Remark 2*: If the non-networked closed-loop system $\dot{x} = f(t, x, ..., x)$ is linear and stable, then the Assumption 2 and 3 always hold with a quadratic choice of $V_1$. It is expected that the assumptions continue to hold with limited nonlinearity in non-networked system. However, the nonlinearity allowed in the plant dynamics is not restricted necessarily by restricting nonlinearity of the non-networked system. An interesting case is when the controllers in augmented system are designed using the feedback linearization method. In this case, the non-networked close loop dynamics $\dot{x} = f(t, x, ..., x)$ is linear but the networked system (1) can be highly nonlinear (because the nonlinear terms are no longer canceled due to the existence of delays).

Before proceeding, we state the following lemma which will be utilized in the remaining of this section.

*Lemma 1*: If the following matrix inequality is satisfied

$$\begin{bmatrix} X_{11} & X_{12} & X_{13} \\ * & X_{22} & X_{23} \\ * & * & X_{33} \end{bmatrix} \geq 0$$

for $X_{ij} \in \mathbb{R}^{n \times n}$ with $1 \leq i, j \leq 3$, then the following integral inequality holds

$$-\int_{t-h(t)}^{t} \dot{x}^T(s) X_{33} \dot{x}(s)$$

$$\leq \int_{t-h(t)}^{T} \begin{bmatrix} x(t) \\ x(t-h(t)) \\ \dot{x}(s) \end{bmatrix}^T \begin{bmatrix} X_{11} & X_{12} & X_{13} \\ * & X_{22} & X_{23} \\ * & * & 0 \end{bmatrix} \begin{bmatrix} x(t) \\ x(t-h(t)) \\ \dot{x}(s) \end{bmatrix}$$

□

The proof is straightforward and thereof omitted.

The following theorem is the main result of this paper which can be regarded as an extended and improved version of the Theorem 1 in [10].

*Theorem 1*: If the Assumptions 1 through 3 are satisfied, then the system (1) is asymptotically stable if there exists diagonal matrices with positive elements $R_k, Y_1, Y_2$ and arbitrary matrices $X_{11}^{(k)}, X_{12}^{(k)}, X_{22}^{(k)}$ for $1 \leq k \leq q$ all belonging to $\mathbb{R}^{n \times n}$ such that the following LMIs are satisfied.

$$\begin{bmatrix} X_{11}^{(k)} & X_{12}^{(k)} & -W^T M_k - Y_1 M_k \\ * & X_{22}^{(k)} & -Y_2 M_k \\ * & * & R_k \end{bmatrix} \geq 0, k \in \{1, ..., q\} \quad (6a)$$

$$\begin{bmatrix} -S + 2Y_1 F + \sum_{k=1}^{q} r_k X_{11}^{(k)} & F^T Y_2 + Y_1 + \sum_{k=1}^{q} r_k X_{12}^{(k)} \\ * & -2Y_2 + \sum_{k=1}^{q} r_k (X_{22}^{(k)} + R_k) \end{bmatrix} < 0 \quad (6b)$$

*Proof:* We use the following Lyapunov-Krasovskii functional which is a modified form of the functional based on the implicit model transformation for linear time-delay systems (see [13]).

$$V = V_1(t, x(t)) + \sum_{k=1}^{q} \int_{t-r_k}^{t} (\theta - t + r_k) \dot{x}^T(\theta) R_k \dot{x}(\theta) d\theta. \quad (7)$$

It is noted that the above functional is more general than the Lyapunov functional used in [10].

The time derivative of $V$ in (7) is calculated as:

$$\dot{V} = \frac{\partial V_1}{\partial t} + \frac{\partial V_1}{\partial t} f\left(t, x, x_{d_1}, \ldots, x_{d_q}\right) +$$

$$\sum_{k=1}^{q} r_k \dot{x}^T R_k \dot{x} - \sum_{k=1}^{q} \int_{t-r_k}^{t} \dot{x}^T(\theta) R_k \dot{x}(\theta) d\theta. \tag{8}$$

Equation (3), in the following with $\Gamma_i$ defined in (5b) can be verified by eliminating the opposite terms form the right hand side.

$$f\left(t, x, x_{d_1}, \ldots, x_{d_q}\right) = f(t, x, \ldots, x) + \sum_{i=k}^{q} [\Gamma_k - \Gamma_{k+1}]. \tag{9}$$

Replacing (9) in (8), results in (10). A vanishing expression is also added at the last line of (10) which is a combination of (9) and (1) multiplied by $(x^T Y_1 + \dot{x} Y_2)$.

$$\dot{V} = \left[\frac{\partial V_1}{\partial t} + \frac{\partial V1}{\partial x} f(t, x, \ldots, x)\right] + \frac{\partial V_1}{\partial x} \sum_{k=1}^{q} [\Gamma_k - \Gamma_{k+1}]$$
$$+ \sum_{k=1}^{q} r_k \dot{x}^T R_k \dot{x} - \sum_{k=1}^{q} \int_{t-r_k}^{t} \dot{x}^T(\theta) R_k \dot{x}(\theta) d\theta$$
$$+ 2(x^T Y_1 + \dot{x}^T Y_2)\left(-\dot{x} + f(t, x, \ldots, x) + \sum_{k=1}^{q} [\Gamma_k - \Gamma_{k+1}]\right). \tag{10}$$

The element-wise absolute values satisfies $a^T M b \leq \bar{a}^T \overline{M} \bar{b}$ for vectors $a, b$ and matrix $M$ with compatible dimensions. Based on this, using (2) and expanding the last expression in (10) we can find an upper bound for $\dot{V}$ as

$$\dot{V} \leq -\bar{x}^T S \bar{x}$$
$$+ \overline{\frac{\partial V_1}{\partial x}} \sum_{k=1}^{q} \overline{[\Gamma_k - \Gamma_{k+1}]} + \sum_{k=1}^{q} r_k \bar{x}^T R_k \bar{x}$$
$$- \sum_{k=1}^{q} \int_{t-r_k}^{t} \bar{x}^T(\theta) R_k \bar{x}(\theta) d\theta$$
$$+ 2\left(\bar{x}^T Y_1 \bar{x} - \bar{x}^T Y_2 \bar{x} + \bar{x}^T Y_1 \overline{f(t, x, \ldots, x)}\right.$$
$$+ \bar{x}^T Y_2 \overline{f(t, x, \ldots, x)}$$
$$+ \bar{x}^T Y_1 \sum_{k=1}^{q} \overline{[\Gamma_k - \Gamma_{k+1}]}$$
$$\left. + \bar{x}^T Y_2 \sum_{k=1}^{q} \overline{[\Gamma_k - \Gamma_{k+1}]}\right). \tag{11}$$

It is noted that we have used $\dot{x}^T Y_2 \dot{x} = \bar{x}^T Y_2 \bar{x}$ which holds due to non-negativeness of all the elements of of $Y_2$ and $\bar{x}$. Substituting relevant terms from the inequalities (3), (4) and (5a) we have

$$\dot{V} \leq -\bar{x}^T S \bar{x} + 2 \sum_{k=1}^{q} \bar{x}^T W^T M_k \overline{(x - x_{d_k})}$$

$$+ \sum_{k=1}^{q} r_k \bar{x}^T R_k \bar{x}$$

$$- \sum_{k=1}^{q} \int_{t-r_k}^{t} \bar{x}^T(\theta) R_i \bar{x}(\theta)$$

$$+ 2 \left( \bar{x}^T Y_1 \tilde{x} - \tilde{x} Y_2 \bar{x} + \bar{x}^T Y_1 F \bar{x} + \tilde{x}^T Y_2 F \bar{x} \right.$$

$$+ \bar{x}^T Y_1 \sum_{k=1}^{q} M_k \overline{(x - x_{d_k})}$$

$$\left. + \tilde{x}^T Y_2 \sum_{k=1}^{q} M_k \overline{(x - x_{d_k})} \right).$$ (12)

The term $\overline{x - x_{d_l}}$ can be calculated and bounded as below

$$x - x_{d_k} = \int_{t-d_k}^{t} \dot{x}(\theta) d\theta \implies$$

$$\overline{x - x_{d_k}} \leq \int_{t-d_k}^{t} \bar{x}(\theta) d\theta \leq \int_{t-r_k}^{t} \bar{x}(\theta) d\theta.$$ (13)

Replacing the above bounding in (12) we obtain

$$\dot{V} \leq -\bar{x}^T S \bar{x} + \sum_{k=1}^{q} r_k \bar{x}^T R_k \bar{x}$$

$$+2(\bar{x}^T Y_1 \tilde{x} + \bar{x}^T Y_1 F \bar{x} - \tilde{x} Y_2 \bar{x} + \tilde{x}^T Y_2 F \bar{x})$$

$$+2 \sum_{k=1}^{q} \int_{t-r_k}^{t} \left( \bar{x}^T Y_1 M_k \bar{x}(\theta) \right.$$
$$+ \bar{x}^T W^T M_k \bar{x}(\theta) + \tilde{x}^T Y_2 M_k \bar{x}(\theta)$$
$$\left. - \bar{x}^T(\theta) R_k \bar{x}(\theta) \right).$$ (14)

Applying Lemma 1 to the matrices

$$\begin{bmatrix} X_{11}^{(k)} & X_{12}^{(k)} & -W^T M_k - Y_1 M_k \\ * & X_{22}^{(k)} & -Y_2 M_k \\ * & * & R_k \end{bmatrix} \geq 0, (1 \leq k \leq q)$$ (15)

with $X_{11}^{(k)}, X_{12}^{(k)}, X_{22}^{(k)} \in \mathbb{R}^{n \times n}$ for $1 \leq k \leq q$, we have

$$- \sum_{k=1}^{q} \int_{t-r_k}^{t} \bar{x}(\theta)^T R_k \bar{x}(\theta)$$

$$\leq \sum_{k=1}^{q} \int_{t-r_k}^{t} \begin{bmatrix} \bar{x} \\ \tilde{x} \\ \bar{x}(\theta) \end{bmatrix}^T \begin{bmatrix} X_{11}^{(k)} & X_{12}^{(k)} & -WM_k - Y_1 M_k \\ * & X_{22}^{(k)} & -Y_2 M_k \\ * & * & 0 \end{bmatrix} \begin{bmatrix} \bar{x} \\ \tilde{x} \\ \bar{x}(\theta) \end{bmatrix} d\theta.$$

Now the integrated terms in (14) can be eliminated by using the above inequalities as below.

$$\dot{V} \leq -\bar{x}^T S \bar{x} + \sum_{k=1}^{q} \bar{\dot{x}}^T R_k \bar{\dot{x}} + 2\bar{x}^T Y_1 \bar{\dot{x}} +$$

$$2\bar{x}^T Y_1 F \bar{x} - 2\bar{\dot{x}}^T Y_2 \bar{\dot{x}} + 2\bar{\dot{x}}^T Y_2 F \bar{x} + \sum_{k=1}^{q} r_k \left( \bar{x}^T X_{11}^{(k)} \bar{x} + \bar{\dot{x}}^T X_{22}^{(k)} \bar{\dot{x}} + 2\bar{x}^T X_{12}^{(k)} \bar{\dot{x}} \right)$$

The above inequality can be written as below:

$$\dot{V} \leq \begin{bmatrix} \bar{x} \\ \bar{\dot{x}} \end{bmatrix}^T \Phi \begin{bmatrix} \bar{x} \\ \bar{\dot{x}} \end{bmatrix}$$

$$\Phi = \begin{bmatrix} -S + 2Y_1 F + \sum_{k=1}^{q} r_i X_{11}^{(k)} & F^T Y_2 + Y_1 + \sum_{k=1}^{q} r_k X_{12}^{(k)} \\ * & -2Y_2 + \sum_{k=1}^{q} r_k \left( X_{22}^{(k)} + R_k \right) \end{bmatrix}$$

If the matrix $\Phi$ is negative definite, then according to the Lyapanov-Krasovskii theorem [12], the system (1) is asymptotically stable which completes the proof. □

## IV. CASE STUDY

Consider a networked robot arm with two degrees of freedom, in which a network cable is passed through the robot manipulator links. The $i$th joint contains an actuator to produce input torque $\tau_i$, and a sensor that measures the angular position of the $i$th joint $q_i$ and its angular velocity $\dot{q}_i$ for $i \in \{1,2\}$. In the $k$th control cycle, the controller polls the $i$th sensor at $t_{k,i}^s$ to calculate $\tau_i$ at $t_k^c$ and sends them to the $i$th actuator at $t_{k,i}^a$, $i \in \{1,2\}$. The length of control cycle is bounded by $T$ such that $t_{k+1,1}^s - t_{k,1}^s < T$ for every $k$. Our aim is to find the maximum value for $T$ such that the stability of the robot arm control system is guaranteed.

The equations of the motion of the robot arm are

$$\begin{aligned} & M_{11}(q_2)\ddot{q}_1 + M_{12}(q_2)\ddot{q}_2 \\ & - N_1(q_1, \dot{q}_1, q_2, \dot{q}_2) = \tau_1 \\ & M_{12}(q_2)\ddot{q}_1 + M_{22}(q_2)\ddot{q}_2 \\ & - N_2(q_1, \dot{q}_1, q_2, \dot{q}_2) = \tau_2 \end{aligned} \quad (16)$$

with the following definitions

$$M_{11}(q_2) = (m_1 + m_2)a_1^2 + m_2 a_2^2 + 2m_2 a_1 a_2 \cos(q_2)$$
$$M_{12}(q_2) = m_2 a_2 [a_2 + a_1 \cos(q_2)]$$
$$M_{22}(q_2) = m_2 a_2^2$$
$$\begin{aligned} N_1(q_1, \dot{q}_1, q_2, \dot{q}_2) = & -m_2 a_1 a_2 (2\dot{q}_1 \dot{q}_2 + \dot{q}_2^2) \sin(q_2) \\ & + (m1 + m2) g a_1 \cos(q_1) \\ & + m_2 g a_2 \cos(q_1 + q_2) \end{aligned}$$

$$N_2(q_1, \dot{q}_1, q_2, \dot{q}_2) = m_2 a_1 a_2 \dot{q}_1 \sin(q_2) + m_2 g a_2 \cos(q_1 + q_2) \tag{17}$$

where $m_1 = 1.5$ Kg, $m_2 = .8$ Kg, $a_1 = 0.5$m, $a_2 = 0.4$m and $g = 9.8$m/sec2 are the weight of link 1, weight of link 2, length of link 1, length of link 2 and acceleration of gravity respectively (see [12]).

In the non-networked case, the robot controller can be designed such that $q_1$ and $q_2$ track the setpoints $q_{d_1}$ and $q_{d_2}$ using the computed torque or the feedback linearization method. Defining the tracking errors as

$$e_i = q_i - q_{d_i} \quad i \in \{1,2\}$$

the closed loop error dynamics for applying the feedback linearization method are selected as

$$\ddot{e}_i + \alpha_i \dot{e}_i + \beta_i e_i = 0 \quad i \in \{1,2\} \tag{18}$$

by choosing $\alpha_i$ and $\beta_i$ appropriately.

In the networked case, the networking effects cause delays in the data path from the sensors to the controller and from the controller to the actuators. Although the closed loop equations in (18) resulting from the feedback linearization method are a linear set of equations, but due to existence of delays the closed loop equations become highly nonlinear in the networked case. Considering the communications to and from the controller mentioned at the beginning of the section, we have four delay values in the system as

$$d_1(t) = d_{s_1} + d_{a_1}, \quad d_2(t) = d_{s_1} + d_{a_2},$$
$$d_3(t) = d_{s_2} + d_{a_1}, \quad d_4(t) = d_{s_2} + d_{a_2}.$$

where $d_{s_i}, i \in \{1,2\}$ is the delay from $i$th sensor to the controller and $d_{a_i}, i \in \{1,2\}$ is the delay from the controller to $i$th actuator. It can be shown that $d_i(t) < 2T$ (see [10]).

For simplicity, we assume that the setpoints $q_{d_i}, i \in \{1,2\}$ are constant such that we can write (18) as

$$\ddot{q}_i + \alpha_i \dot{q}_i + \beta_i(q_i - q_{d_i}) = 0 \quad i \in \{1,2\}$$

Using the feedback linearization method, the feedback law for the non-networked case is obtained by replacing $\ddot{q}_i, i \in \{1,2\}$ in (16) from the above equation as below.

$$\tau_1 = -M_{11}(q_2)[\alpha_1 \dot{q}_1 + \beta_1(q_1 - q_{d_1})] - M_{12}(q_2)[\alpha_2 \dot{q}_2 + \beta_2(q_2 - q_{d_2})] - N_1(q_1, \dot{q}_1, q_2, \dot{q}_2)$$

$$\tau_2 = -M_{12}(q_2)[\alpha_1 \dot{q}_1 + \beta_1(q_1 - q_{d_1})] - M_{22}(q_2)[\alpha_2 \dot{q}_2 + \beta_2(q_2 - q_{d_2})] - N_2(q_1, \dot{q}_1, q_2, \dot{q}_2)$$

In the networked case the time varying delays $d_i, 1 \leq i \leq 4$ are inserted into the above equations as

$$\tau_1 = -M_{11}(q_{2,3})[\alpha_1 \dot{q}_{1,1} + \beta_1(q_{1,1} - q_{d_1})]$$
$$- M_{12}(q_{2,3})[\alpha_2 \dot{q}_{2,3} + \beta_2(q_{2,3} - q_{d_2})]$$
$$- N_1(q_{1,1}, \dot{q}_{1,1}, q_{2,3}, \dot{q}_{2,3})$$

$$\tau_2 = -M_{12}(q_{2,4})[\alpha_1 \dot{q}_{1,2} + \beta_1(q_{1,2} - q_{d_1})]$$
$$- M_{22}(q_{2,4})[\alpha_{2,4} \dot{q}_2 + \beta_2(q_{2,4} - q_{d_2})]$$
$$- N_2(q_{1,2}, \dot{q}_{1,2}, q_{2,4}, \dot{q}_{2,4}) \tag{19}$$

where $q_i\bigl(t - d_j(t)\bigr), i \in \{1,2\}, k \in \{1,2,3,4\}$ is written as $q_{i,k}$ for compactness.

By selecting the state vector as $x = [q_1 \; \dot{q}_1 \; q_2 \; \dot{q}_2]^T$ the closed loop equations for the networked case are obtained by replacing $\tau_i, i \in \{1,2\}$ in (16) from (19). As mentioned in Remark 2, the closed loop equations for the networked case are delayed and highly nonlinear compared to the non-networked closed loop equations in (18).

Non-networked error dynamics are selected as $\alpha_1 = \alpha_2 = 2.55$ and $\beta_1 = \beta_2 = 3.16$ similar to [10]. According to the Assumption 2, for applying the Theorem 1 we first need to find a Lyapunov function $V_1$ for the closed loop dynamics. Due to using feedback linearization the non-networked closed loop dynamics are in the form of $\dot{x} = Ax$ with $A \in \mathbb{R}^{n_x \times n_x}$ being stable. Therefore, $F$ in (4) is easily obtained as

$$F = \bar{A} = \begin{bmatrix} 0 & 1 & 0 & 0 \\ 3.16 & 2.55 & 0 & 0 \\ 0 & 0 & 0 & 1 \\ 0 & 0 & 3.16 & 2.55 \end{bmatrix}$$

As a replacement for the suggestion in Remark 4.1 in [10], it is suggested to select $V_1$ in as $V_1(x) = x^T P x$ with $P$ obtained from maximizing $\alpha > 0$ over the set of LMIs $P > 0$, $P \le I$, $A^T P + PA \le -Q$ and $Q > \alpha I$ to maximize the the smallest eigenvalue of $Q$ for a bounded $P$. By solving these LMIs and considering the fact that $\partial V_1/\partial x = 2Px$, we can calculate $W$ in (3) as

$$W = \bar{P} = \begin{bmatrix} 0.9796 & 0.1271 & 0 & 0 \\ 0.1271 & 0.2074 & 0 & 0 \\ 0 & 0 & 0.9796 & 0.1271 \\ 0 & 0 & 0.1271 & 0.2074 \end{bmatrix}$$

If we restrict the off-diagonal elements of $Q$ to be non-positive (by some additional LMIs), then we can write $x^T Q x \ge \bar{x}^T Q \bar{x}$ and the matrix $Q$ can be a choice for $S$ in (2). Based on this, we obtain $S$ as

$$S = \begin{bmatrix} 0.8035 & -0.0000 & 0 & 0 \\ -0.0000 & 0.8035 & 0 & 0 \\ 0 & 0 & 0.8035 & -0.0000 \\ 0 & 0 & -0.0000 & 0.8035 \end{bmatrix}$$

The Assumption 2 is satisfied with the choices for the matrices *F, W, S* given above. For calculation of the matrices $M_k$, $1 \leq k \leq 4$ in equation (5) for satisfying Assumption 3, it is noticed that all of the elements on the first and third rows of $M_k$, $1 \leq k \leq 4$ are zero. Because, the first and second equations of (1) in the case of robot (which is obtained from equations 16 and 19) are simply written as $[\dot{x}]_1 = [x_2]$ and $[\dot{x}]_3 = [x_4]$ without any delayed term. Also, the third and forth columns of $M_1$, $M_2$ and the first and second columns of $M_3$, $M_4$ are zero. Because, the equations in (19) and thus the elements of the function $f$ in (1) do not depend on some of the delayed state variables. Based on these facts and by thorough search, the matrices $M_k$, $1 \leq k \leq 4$ are computed as

$$M_1 = \begin{bmatrix} 0 & 0 & 0 & 0 \\ 29.7955 & 13.6 & 0 & 0 \\ 0 & 0 & 0 & 0 \\ 67.0399 & 30.6 & 0 & 0 \end{bmatrix}, M_2 = \begin{bmatrix} 0 & 0 & 0 & 0 \\ 22 & 22.1 & 0 & 0 \\ 0 & 0 & 0 & 0 \\ 76.6358 & 75.6 & 0 & 0 \end{bmatrix}$$

$$M_3 = \begin{bmatrix} 0 & 0 & 0 & 0 \\ 0 & 0 & 137.3 & 11.8 \\ 0 & 0 & 0 & 0 \\ 0 & 0 & 308.8420 & 20.4 \end{bmatrix}, M_3 = \begin{bmatrix} 0 & 0 & 0 & 0 \\ 0 & 0 & 104.9 & 2 \\ 0 & 0 & 0 & 0 \\ 0 & 0 & 372.7103 & 7 \end{bmatrix}$$

By applying the Theorem 1 the bound on the control cycle that guarantees the robot arm NCS stability is obtained as $0.7915\ ms$. This result is compared with the results of applying the methods in [10] and [2] for the analysis of the robot arm NCS of this section in Table I (both of the values compared to our result are from [10]). The comparison shows that a considerable improvement is achieved in the analysis by applying the new stability condition.

TABLE I.
COMPARISON OF DIFFERENT ANALYSIS METHODS

| Method | Bound on *T* |
|---|---|
| Theorem 1 | 0.7915 msec |
| The method in [10] | 0.15 msec |
| The method in [2] | 0.033 msec |

## V. Conclusion

In this work, a sufficient condition was presented for the stability analysis of nonlinear networked control systems. The basic idea is to make improvements to a previous result on applying the Lyapunov Krasovskii method to nonlinear NCS's. Application of the obtained result to robot arm NCS reveals that the new stability condition is much more efficient compared to the previous works.